\documentclass[aps,twocolumn,amsmath,letterpaper]{revtex4}
\usepackage{amssymb}
\usepackage{graphicx}
\usepackage{array}
\usepackage{hhline}
\usepackage{longtable}

\renewcommand{\thetable}{\Roman{table}} \thetable

\newcommand{\Jz}{J_{\!z}}
\newcommand{\Jxy}{J_{\!xy}}

\begin{document}

    \title{Excitation Spectrum Gap and Spin-Wave Velocity of
    XXZ Heisenberg Chains:

     Global Renormalization-Group Calculation}

    \author{Ozan S. Sar\i yer$^{1}$, A. Nihat Berker$^{2-4}$, and Michael Hinczewski$^{4}$}
    \affiliation{$^1$Department of Physics, Istanbul Technical University, Maslak 34469, Istanbul, Turkey,}
    \affiliation{$^2$College of Sciences and Arts, Ko\c{c} University, Sar\i yer 34450, Istanbul, Turkey,}
\affiliation{$^3$Department of Physics, Massachusetts Institute of
Technology, Cambridge, Massachusetts 02139, U.S.A.,}
    \affiliation{$^4$Feza G\"ursey Research Institute, T\"UB\.ITAK - Bosphorus University, \c{C}engelk\"oy 34684, Istanbul, Turkey}

    \begin{abstract}
The anisotropic XXZ spin-$\frac{1}{2}$ Heisenberg chain is studied
using renormalization-group theory.  The specific heats and
nearest-neighbor spin-spin correlations are calculated thoughout the
entire temperature and anisotropy ranges in both ferromagnetic and
antiferromagnetic regions, obtaining a global description and
quantitative results.  We obtain, for all anisotropies, the
antiferromagnetic spin-liquid spin-wave velocity and the Isinglike
ferromagnetic excitation spectrum gap, exhibiting the spin-wave to
spinon crossover.  A number of characteristics of purely quantum
nature are found: The in-plane interaction
$s_{i}^{x}s_{j}^{x}+s_{i}^{y}s_{j}^{y}$ induces an antiferromagnetic
correlation in the out-of-plane $s_{i}^{z}$ component, at higher
temperatures in the antiferromagnetic XXZ chain, dominantly at low
temperatures in the ferromagnetic XXZ chain, and, in-between, at all
temperatures in the XY chain.  We find that the converse effect also
occurs in the antiferromagnetic XXZ chain: an antiferromagnetic
$s_{i}^{z}s_{j}^{z}$ interaction induces a correlation in the
$s_{i}^{xy}$ component.  As another purely quantum effect, (i) in
the antiferromagnet, the value of the specific heat peak is
insensitive to anisotropy and the temperature of the specific heat
peak decreases from the isotropic (Heisenberg) with introduction of
either type (Ising or XY) anisotropy; (ii) in complete contrast, in
the ferromagnet, the value and temperature of the specific heat peak
increase with either type of anisotropy.

    \end{abstract}

    \maketitle
    \def\s{\rule{0in}{0.28in}}

    \section{Introduction}

    \setlength{\LTcapwidth}{\columnwidth}

The quantum Heisenberg chain, including the possibility of
spin-space anisotropy, is the simplest nontrivial quantum spin
system and has thus been widely studied since the very beginning of
the spin concept in quantum mechanics \cite{Bloch,Bethe,Hulthen}.
Interest in this model continued \cite{Lieb, Katsura, BF,
Inawashiro, Yang, Johnsonb, Johnson} and redoubled with the
exposition of its richly varied low-temperature behavior
\cite{Haldane1980, Haldane1982, Haldane1983} and of its relevance to
high-temperature superconductivity \cite{Bednorz, Manousakis,
Keimer, Greven, Birge}. It has become clear that antiferromagnetism
and superconductivity are firmly related to each other, adjoining
and overlapping each other.

A large variety of theoretical tools have been employed in the study
of the various isotropic and anisotropic regimes of the quantum
Heisenberg chain, including finite-systems extrapolation \cite{BF,
Narayanan}, linked-cluster \cite{Inawashiro} and dimer-cluster
\cite{Karbach} expansions, quantum decimation \cite{Xi-Yao},
decoupled Green's functions \cite{Zhang}, quantum transfer matrix
\cite{Fabricius, Klumper}, high-temperature series expansion
\cite{Rojas}, and numerical evaluation of multiple integrals
\cite{Bortz}.  The XXZ Heisenberg chain retains high current
interest as a theoretical model \cite{Damerau, Hu} with direct
experimental relevance \cite{Thanos}.

In the present paper, a position-space renormalization-group method
introduced by Suzuki and Takano \cite{Suzuki,Takano} for $d=2$
dimensions and already applied to a number of $d>1$ systems
\cite{Suzuki, Takano, Falicov, Tomczak, TomRich1, TomRich2,
Hinczewski} is used to compute the spin-spin correlations and the
specific heat of the $d=1$ anisotropic quantum XXZ Heisenberg model,
easily resulting in a global description and detailed quantitative
information for the entire temperature and anisotropy ranges
including the ferromagnetic and antiferromagnetic, the spin-liquid
and Isinglike regions.  We obtain, for all anisotropies, the
antiferromagnetic spin-liquid spin-wave velocity and the Isinglike
ferromagnetic excitation spectrum gap, exhibiting the spin-wave to
spinon crossover.  A number of other characteristics of purely
quantum nature are found: The in-plane interaction
$s_{i}^{x}s_{j}^{x}+s_{i}^{y}s_{j}^{y}$ induces an antiferromagnetic
correlation in the out-of-plane $s_{i}^{z}$ component, at higher
temperatures in the antiferromagnetic XXZ chain, dominantly at low
temperatures in the ferromagnetic XXZ chain, and, in-between, at all
temperatures in the XY chain.  We find that the converse effect also
occurs in the antiferromagnetic XXZ chain: an antiferromagnetic
$s_{i}^{z}s_{j}^{z}$ interaction induces a correlation in the
$s_{i}^{xy}$ component.  As another purely quantum effect, (i) in
the antiferromagnet, the value of the specific heat peak is
insensitive to anisotropy and the temperature of the specific heat
peak decreases from the isotropic (Heisenberg) with introduction of
either type (Ising or XY) anisotropy; (ii) in complete contrast, in
the ferromagnet, the value of the specific heat peak is strongly
dependent on anisotropy and the temperature of the specific heat
peak increases with either type of anisotropy. This purely quantum
effect is a precursor to different phase transition temperatures in
three dimensions \cite{Rushbrooke, Oitmaa, Falicov, Hinczewski}. Our
calculational method is relatively simple, readily yields global
results, and is overall quantitatively successful.

    \section{The Anisotropic Quantum Heisenberg Model and the Renormalization-Group Method}

    \subsection{The Anisotropic Quantum Heisenberg Model}

The spin-$\frac{1}{2}$ anisotropic Heisenberg model (XXZ model) is
defined by the dimensionless Hamiltonian

    \begin{align}
        -\beta \mathcal{H}=\sum_{\langle ij \rangle}\left\{\left[
        J_{xy}\left(s_{i}^{x}s_{j}^{x}+s_{i}^{y}s_{j}^{y}\right)+J_{z}s_{i}^{z}s_{j}^{z}\right]+G\right\},\label{eq:1}
    \end{align}
where $\beta=1/k_{B}T$ and $\langle ij \rangle$ denotes summation
over nearest-neighbor pairs of sites.  Here the $s_{i}^u$ are the
quantum mechanical Pauli spin operators at site $i$.  The additive
constant $G$ is generated by the renormalization-group
transformation and is used in the calculation of thermodynamic
functions.  The anisotropy coefficient is $R = J_{z}/J_{xy}$.  The
model reduces to the isotropic Heisenberg model (XXX model) for
$|R|=1$, to the XY model for $R=0$, and to the Ising model for
$|R|\rightarrow\infty$.

    \subsection{Renormalization-Group Recursion Relations}

    The Hamiltonian in Eq.(\ref{eq:1}) can be rewritten as

    \begin{equation}
        -\beta \mathcal{H}=\sum_{i}\left\{ -\beta
        \mathcal{H}(i,i+1)\right\}. \label{eq:2}
    \end{equation}

\noindent where $\beta \mathcal{H}(i,i+1)$ is a Hamiltonian
involving sites $i$ and $i+1$ only. The renormalization-group
procedure, which eliminates half of the degrees of freedom and keeps
the partition function unchanged, is done approximately
\cite{Suzuki,Takano}:

    \begin{align}
        \label{eq:3}
            \mbox{Tr}_{\mbox{\tiny odd}}e^{-\beta
            \mathcal{H}} =&\mbox{Tr}_{\mbox{\tiny
            odd}}e^{\sum_{i}\left\{ -\beta \mathcal{H}(i,i+1)\right\} }\\
\nonumber =&\mbox{Tr}_{\mbox{\tiny odd}} e^{\sum_{i}^{\mbox{\tiny
            odd}}\left\{ -\beta \mathcal{H}(i-1,i)-\beta \mathcal{H}(i,i+1) \right\} }\\
\nonumber\simeq& \prod_{i}^{\mbox{\tiny odd}}\mbox{Tr}_{i}e^{\left\{
            -\beta \mathcal{H}(i-1,i)-\beta \mathcal{H}(i,i+1)\right\} }\\
\nonumber=\prod_{i}^{\mbox{\tiny
            odd}}e^{-\beta ^{\prime }\mathcal{H}^{\prime }(i-1,i+1)}&
            \simeq e^{\sum_{i}^{\mbox{\tiny odd}}\left\{ -\beta ^{\prime
            }\mathcal{H}^{\prime }(i-1,i+1)\right\} } =e^{-\beta ^{\prime
            }\mathcal{H}^{\prime }}.
        \end{align}
Here and throughout this paper, the primes are used for the
renormalized system.  Thus, at each successive length scale, we
ignore the non-commutativity of the operators beyond three
consecutive sites, in the two steps indicated by $\simeq$ in the
above equation.  Since the approximations are applied in opposite
directions, one can expect some mutual compensation.  Earlier
studies \cite{Suzuki,Takano, Tomczak, TomRich1, TomRich2} have been
successful in obtaining finite-temperature behavior on a variety of
quantum systems.

\begin{table}[h]
        \begin{tabular}{|c|c|c|c|}
            \hline
            $p$ & $s$ & $m_s$ & Two-site basis eigenstates${{}}^{{^{{^{{}}}}}}$\\
            \hline
            $+$ & $1$ & $1$ &$|\phi_{1}\rangle=|\uparrow\uparrow\rangle$ ${{}}^{{^{{^{{}}}}}}$ \\
            \cline{3-4}
            $ $ & $ $ & $0$ &$|\phi_{2}\rangle=\frac{1}{\sqrt{2}}\{|\uparrow\downarrow\rangle+|\downarrow\uparrow\rangle\}$ ${{}}^{{^{{^{{}}}}}}$\\
            \hline
            $-$ & $0$ & $0$ &$|\phi_{4}\rangle=\frac{1}{\sqrt{2}}\{|\uparrow\downarrow\rangle-|\downarrow\uparrow\rangle\}$ ${{}}^{{^{{^{{}}}}}}$\\
            \hline
        \end{tabular}
        \caption{The two-site basis eigenstates that appear in
        Eq.(\ref{eq:6}). These are the well-known singlet and triplet
        states. The state $|\phi_{3}\rangle$ is obtained by spin reversal
        from $|\phi_{1}\rangle$, with the same eigenvalue.}
        \label{tab:1}
    \end{table}

    The transformation above is summarized by

    \begin{equation}
        \label{eq:4} e^{-\beta ^{\prime }\mathcal{H}^{\prime }(i,k)}
        =\mbox{Tr}_{\!\tiny j\,}e^{\left\{ -\beta \mathcal{H}(i,j)-\beta
        \mathcal{H}(j,k)\right\}},
    \end{equation}
    where $i, j, k$ are three successive sites. The operator
    $-\beta^{\prime } \mathcal{H}^{\prime }(i,k)$ acts on two-site states, while the operator
    $-\beta \mathcal{H}(i,j)-\beta \mathcal{H}(j,k)$ acts on three-site
    states, so that we can rewrite Eq.(\ref{eq:4}) in the matrix form,
\begin{multline}
        \langle u_{i}v_{k}|e^{-\beta ^{\prime }\mathcal{H}^{\prime
        }(i,k)}|\bar{u}_{i}^{{}}\bar{v}_{k}^{{}}\rangle = \label{eq:5}\\
        \sum_{w_{j}}\langle u_{i}\,w_{j}\,v_{k}|e^{-\beta
        \mathcal{H}(i,j)-\beta
        \mathcal{H}(j,k)}|\bar{u}_{i}\,w_{j}\,\bar{v}_{k}^{{}}\rangle \:,
    \end{multline}
where state variables $u,v,w,\bar{u},\bar{v}$ can take spin-up or
spin-down values at each site.  The unrenormalized $8\times8$ matrix
on the right-hand side is contracted into the renormalized
$4\times4$ matrix on the left-hand side of Eq.(\ref{eq:5}). We use
two-site basis states vectors $\{|\phi_{p}\rangle \}$ and three-site
basis states vectors $\{|\psi_{q}\rangle \}$ to diagonalize the
matrices in Eq.(\ref{eq:5}). The states $\{|\phi_{p}\rangle \}$,
given in Table \ref{tab:1}, are eigenstates of parity, total spin
magnitude, and total spin z-component.  These $\{|\phi_{p}\rangle
\}$ diagonalize the renormalized matrix, with eigenvalues

\begin{align}
\nonumber \Lambda_{1} =& \frac{1}{4} J_{\!z}^\prime + G ^\prime,
\qquad
\Lambda_{2} = + \frac{1}{2} J_{\!xy}^\prime -\frac{1}{4} J_{\!z}^\prime + G ^\prime,\\
&\Lambda_{4} = - \frac{1}{2} J_{\!zxy}^\prime  -\frac{1}{4}
J_{\!z}^\prime + G ^\prime.
\end{align}
The states $\{|\psi_{q}\rangle \}$, given in Table \ref{tab:2}, are
eigenstates of parity and total spin z-component. The
$\{|\psi_{p}\rangle \}$ diagonalize the unrenormalized matrix, with
eigenvalues
\begin{align}
\label{eq:8} \lambda_{1} &= \frac{1}{2} J_{\!z} + 2 G, \qquad \lambda_{4} = 2 G, \\
\nonumber \lambda_{2} &= -\frac{1}{4}\left(J_{\!z}+\sqrt{8 J_{\!xy}^2 + J_{\!z}^2}\right) + 2 G, \\
\nonumber \lambda_{3} &= -\frac{1}{4}\left(J_{\!z}-\sqrt{8J_{\!xy}^2
+ J_{\!z}^2}\right) + 2 G.
\end{align}

        \begin{table}[h]
        \begin{tabular}{|c|c|c|}
            \hline
            $p$ & $m_s$ & Three-site basis eigenstates${{}}^{{^{{^{{}}}}}}$\\
            \hline
            $+$ & $3/2$ &$|\psi_{1}\rangle=|\uparrow\uparrow\uparrow\rangle$ ${{}}^{{^{{^{{}}}}}}$ \\
            \cline{2-3}
            $ $ & $1/2$ &$|\psi_{2}\rangle=\mu\{|\uparrow\uparrow\downarrow\rangle+\sigma|\uparrow\downarrow\uparrow\rangle+|\downarrow\uparrow\uparrow\rangle\}$ ${{}}^{{^{{^{{}}}}}}$\\
            \cline{3-3}
            $ $ & $ $ &$|\psi_{3}\rangle=\nu\{-|\uparrow\uparrow\downarrow\rangle+\tau|\uparrow\downarrow\uparrow\rangle-|\downarrow\uparrow\uparrow\rangle\}$ ${{}}^{{^{{^{{}}}}}}$\\
            \hline
            $-$ & $1/2$ &$|\psi_{4}\rangle=\frac{1}{\sqrt{2}}\{|\uparrow\uparrow\downarrow\rangle-|\downarrow\uparrow\uparrow\rangle\}$ ${{}}^{{^{{^{{}}}}}}$\\
            \hline
        \end{tabular}
\caption{The three-site basis eigenstates that appear in
Eq.(\ref{eq:6}) with coefficients $\sigma=(-J_z+\sqrt{8 J_{xy}^2 +
J_z^2})/2 J_{xy}, \tau=(J_z+\sqrt{8 J_{xy}^2 + J_z^2})/2J_{xy}$ and
normalization factors $\mu$, $\nu$. The states $|\psi_{5-8}\rangle$
are obtained by spin reversal from $|\psi_{1-4}\rangle$, with the
same respective eigenvalues. } \label{tab:2}
\end{table}

\noindent With these eigenstates, Eq.(\ref{eq:5}) is rewritten as

    \begin{multline}
        \gamma_{p}\equiv \langle \phi _{p}|e^{-\beta ^{\prime }\mathcal{H}^{\prime
        }(i,k)}|\phi _p\rangle = \sum_{\substack{u,v,\bar{u},\\ \bar{v},w,q}}
        \langle\phi _p|u_iv_k\rangle \langle
        u_iw_jv_k|\psi_q\rangle \cdot  \label{eq:6}\\
         \langle \psi _q|e^{-\beta
        \mathcal{H}(i,j)-\beta
        \mathcal{H}(j,k)}|\psi _{q}\rangle
        \langle \psi_{q}|\bar{u}_iw_j\bar{v}_k\rangle \langle
        \bar{u}_i\bar{v}_k|\phi _{{p}}\rangle\:.
    \end{multline}

\noindent Thus, there are three independent $\gamma_p$ that
determine the renormalized Hamiltonian and, therefore, three
renormalized interactions in the Hamiltonian closed under
renormalization-group transformation, Eq.(\ref{eq:1}). These
$\gamma_{p}$ are

    \begin{align}
\nonumber\gamma_{1} =
e^{\frac{1}{4}J_{\!z}^\prime+G^\prime}=e^{2G-\frac{1}{4}J_{\!z}}
&\Bigg[e^{\frac{3}{4} J_{\!z}}+\cosh\left(\frac{1}{4}\sqrt{8 J_{\!xy}^2+J_{\!z}^2}\right)\\
\nonumber &-\frac{J_{\!z} \sinh\left(\frac{1}{4}\sqrt{8
J_{\!xy}^2+J_{\!z}^2}\right)}{\sqrt{8J_{\!xy}^2+J_{\!z}^2}}\Bigg]\:,\\
\nonumber
\gamma_{2}=e^{\frac{1}{2}J_{\!xy}^\prime-\frac{1}{4}J_{\!z}^\prime+G^\prime}=&2e^{2G-\frac{1}{4}J_{\!z}}
\Bigg[\cosh\left(\frac{1}{4}\sqrt{8
J_{\!xy}^2+J_{\!z}^2}\right)\\
\nonumber&+\frac{J_{\!z} \sinh\left(\frac{1}{4}\sqrt{8
J_{\!xy}^2+J_{\!z}^2}\right)}{\sqrt{8
J_{\!xy}^2+J_{\!z}^2}}\Bigg]\:,\\
\gamma_{4}=e^{-\frac{1}{2} J_{\!xy}^\prime-\frac{1}{4}
J_{\!z}^\prime+G^\prime}&=2e^{2G}\:,
    \end{align}
which yield the recursion relations

    \begin{align}
\label{eq:13}
J_{\!xy}^\prime=\ln\left(\frac{\gamma_{2}}{\gamma_{4}}\right),
 J_{\!z}^\prime=\ln\left(\frac{\gamma_{1}^2}{\gamma_{2}\gamma_{4}}\right),
 G^\prime=\frac{1}{4}\ln\left(\gamma_{1}^2\gamma_{2}\gamma_{4}\right).
    \end{align}
As expected, $\Jxy^\prime$ and $\Jz^\prime$ are independent of the
additive constant $G$ and the derivative
$\partial_{G}G^\prime=b^d=2$, where $b=2$ is the rescaling factor
and $d=1$ is the dimensionality of the lattice.

For $\Jxy=\Jz$, the recursion relations reduce to the
spin-$\frac{1}{2}$ isotropic Heisenberg (XXX) model recursion
relations, while for $\Jxy=0$ they reduce to spin-$\frac{1}{2}$
Ising model recursion relations.  The $\Jz=0$ subspace (XY model) is
not (and need not be) closed under these recursion relations
\cite{Suzuki,Takano}:  The renormalization-group transformation
induces a positive $\Jz$ value, but the spin-space easy-plane aspect
is maintained.

In addition, there exists a mirror symmetry along the $\Jz$-axis, so
that
$\Jxy^\prime\left(-\Jxy,\Jz\right)=\Jxy^\prime\left(\Jxy,\Jz\right)$
and
$\Jz^\prime\left(-\Jxy,\Jz\right)=\Jz^\prime\left(\Jxy,\Jz\right)$.
The thermodynamics of the system remains unchanged under flipping
the interactions of the $x$ and $y$ spin components, since the
renormalization-group trajectories do not change.  In fact, this is
part of a more general symmetry of the XYZ model, where flipping the
signs of any two interactions leaves the spectrum unchanged
\cite{Yang}.  Therefore, with no loss of generality, we take $\Jxy
> 0$. Independent of the sign of $\Jxy$,
$\Jz>0$ gives the ferromagnetic model and $\Jz<0$ gives the
antiferromagnetic model.

    \subsection{Calculation of Densities and Response Functions by the Recursion-Matrix Method}

Just as the interaction constants of two consecutive points along
the renormalization-group trajectory are related by the recursion
relations, the densities are connected by a recursion matrix
$\hat{T}$, which is composed of derivatives of the recursion
relations. For our Hamiltonian, the recursion matrix and density
vector $\vec{M}$ are

    \begin{equation}
        \nonumber
        \begin{split}
            \hat{T}=\left(
            \begin{array}{ccc}
                \frac{\partial G^\prime}{\partial G}    & \frac{\partial G^\prime}{\partial \Jxy}    & \frac{\partial G^\prime}{\partial \Jz}    \\
                0 & \frac{\partial \Jxy^\prime}{\partial \Jxy} & \frac{\partial \Jxy^\prime}{\partial \Jz} \\
                0 & \frac{\partial \Jz^\prime}{\partial \Jxy}  & \frac{\partial \Jxy^\prime}{\partial \Jz} \\
            \end{array}\right)\:, \qquad\qquad\quad \\
        \end{split}
    \end{equation}
    \begin{equation}\label{eq:19}
        \vec{M}=\left(
        \begin{array}{ccc}
            1 & 2\left\langle s_{i}^{xy}s_{j}^{xy}\right\rangle & \left\langle s_{i}^{z}s_{j}^{z}\right\rangle \\
        \end{array}\right)\:.\qquad\qquad
    \end{equation}
These are densities $M_{\alpha}$ associated with each
    interaction $K_{\alpha}$,

    \begin{equation}\label{eq:20}
        M_{\alpha}=\frac{1}{N_{\alpha}}\frac{\partial \ln Z}{\partial K_{\alpha}}\:,
    \end{equation}
    where $N_{\alpha}$ is the number of $\alpha$-type interactions and $Z$ is the partition function for the system,
    which can be expressed both via the unrenormalized interaction constants as $Z(\vec{K})$ or via the renormalized
    interaction constants as $Z(\vec{K^\prime})$. By using these two equivalent forms, one can formulate the density recursion relation~\cite{McKay}

    \begin{equation}\label{eq:21}
        M_{\alpha}=b^{-d}\sum_{\beta}M_{\beta}^\prime
        T_{\beta\alpha}\:,\qquad T_{\beta\alpha}\equiv\frac{N_{\beta}}{N_{\alpha}}\frac{\partial K_{\beta}^\prime}{\partial
        K_{\alpha}}\:.
    \end{equation}
    Since the interaction constants, under renormalization-group transformation, stay the same at fixed points
    such as critical fixed points or sinks, the above Eq.(\ref{eq:21}) takes the
    form of a solvable eigenvalue equation,

    \begin{equation}\label{eq:22}
        b^d\vec{M}^*=\vec{M}^*\cdot\hat{T}\:,
    \end{equation}
    at fixed points, where $\vec{M}=\vec{M^\prime}=\vec{M}^*$. The fixed point densities are the
    components of the left eigenvector of the recursion matrix with
    left eigenvalue $b^d$~\cite{McKay}.  At ordinary points, Eq.(\ref{eq:21}) is iterated until a
    sink point is reached under successive renormalization-group transformations. In algebraic form, this means

    \begin{equation}\label{eq:23}
        \vec{M}^{(0)}=b^{-nd}\vec{M}^{(n)}\hat{T}^{(n)}\hat{T}^{(n-1)}\cdots\hat{T}^{(1)}\:,
    \end{equation}
    where the upper indices indicate the number of iteration (transformation), with $\vec{M}^{(n)}\simeq \vec{M}^*$.

This method is applied on our model Hamiltonian. The sink of the
system is at infinite temperature $\Jxy^*=\Jz^*=0$ for all initial
conditions $(\Jxy,\Jz)$.

Response functions are calculated by differentiation of densities.
For example, the internal energy is $U=-2\left\langle
s_{i}^{xy}s_{j}^{xy}\right\rangle-R\left\langle
s_{i}^{z}s_{j}^{z}\right\rangle$, employing $T=1/\Jxy$, and
$U=-2\left\langle s_{i}^{xy}s_{j}^{xy}\right\rangle/R-\left\langle
s_{i}^{z}s_{j}^{z}\right\rangle$, employing $T=1/|\Jz|$.  The
specific heat $C=\partial_{T}U$ follows from the chain rule,

    \begin{multline}\label{eq:24}
        C=\Jxy^2\frac{\partial\left(2\left\langle
        s_{i}^{xy}s_{j}^{xy}\right\rangle+R\left\langle
        s_{i}^{z}s_{j}^{z}\right\rangle\right)}{\partial\Jxy},\quad\: \textrm{for } T=1/\Jxy,\\
        C=\Jz^2\frac{\partial\left(2\left\langle
        s_{i}^{xy}s_{j}^{xy}\right\rangle/R+\left\langle
        s_{i}^{z}s_{j}^{z}\right\rangle\right)}{\partial|\Jz|},\quad \textrm{for } T=1/|\Jz|.
    \end{multline}

    \section{Correlations Scanned with Respect to Anisotropy}

\begin{figure}[h]
\centering
\includegraphics*[scale=1]{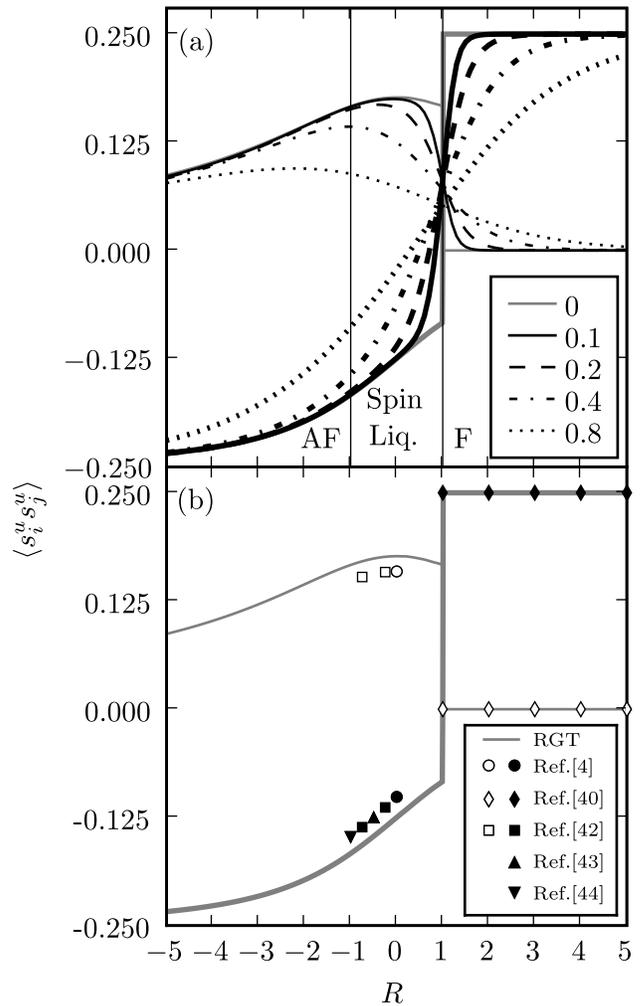}
\caption{(a) Calculated nearest-neighbor spin-spin correlations
$\left\langle s_{i}^{z}s_{j}^{z}\right\rangle$ (thick curves from
lower left) and $\left\langle s_{i}^{xy}s_{j}^{xy}\right\rangle$
(thin curves from upper left) as a function of anisotropy
coefficient $R$ for temperatures $1/\Jxy=0,0.1,0.2,0.4,0.8$. (b)
Calculated zero-temperature nearest-neighbor spin-spin correlations
(thin and thick curves, as in the upper panel) compared with the
exact points of Ref.\cite{Kato, Kitanine, Lieb, Sato, Takahashi}
shown with filled and open symbols for $\left\langle
s_{i}^{z}s_{j}^{z}\right\rangle$ and $\left\langle
s_{i}^{xy}s_{j}^{xy}\right\rangle$ respectively. At $R=1$, the
calculated $\left\langle s_{i}^{z}s_{j}^{z}\right\rangle$
discontinuously goes from antiferromagnetic to the exact result of
0.25 \cite{Takahashi} of saturated ferromagnetism and the calculated
$\left\langle s_{i}^{xy}s_{j}^{xy}\right\rangle$ discontinuously
goes from ferromagnetic to the exact result of constant zero
\cite{Takahashi}.} \label{fig:1}
\end{figure}

The ground-state and excitation properties of the XXZ model offer a
variety of behaviors \cite{Haldane1980, Haldane1982, Dmitriev,
Takahashi}:  The antiferromagnetic model with $R<-1$ is Isinglike
and the ground state has N\'{e}el long-range order along the $z$
spin component with a gap in the excitation spectrum. For $-1\leq
R\leq1$, the system is a "spin liquid", with a gapless spectrum and
power-law decay of correlations at zero temperature.  The
ferromagnetic model with $R>1$ is also Isinglike, the ground state
is ferromagnetic along the $z$ spin component, with an excitation
gap.

Our calculated $\left\langle s_{i}^{z}s_{j}^{z}\right\rangle$ and
$\left\langle s_{i}^{xy}s_{j}^{xy}\right\rangle \equiv \left\langle
s_{i}^{x}s_{j}^{x}\right\rangle = \left\langle
s_{i}^{y}s_{j}^{y}\right\rangle$ nearest-neighbor spin-spin
correlations for the whole range of the anisotropy coefficient $R$
are shown in Fig.\ref{fig:1}, for various temperatures.  The $xy$
correlation is always non-negative.  Recall that we use $\Jxy
> 0$ with no loss of generality. In the Isinglike antiferromagnetic
($R < -1$) region, the $z$ correlation is expectedly
antiferromagnetic. As the $\left\langle
s_{i}^{z}s_{j}^{z}\right\rangle$ correlation saturates for large
$|R|$, the transverse $\left\langle
s_{i}^{xy}s_{j}^{xy}\right\rangle$ correlation is somewhat depleted.
In the Isinglike ferromagnetic ($R>1$) region, the $\left\langle
s_{i}^{z}s_{j}^{z}\right\rangle$ correlation is ferromagnetic,
saturates quickly as the $\left\langle
s_{i}^{xy}s_{j}^{xy}\right\rangle$ correlation quickly goes to zero.
In the spin-liquid ($|R|<1$) region, the $\left\langle
s_{i}^{z}s_{j}^{z}\right\rangle$ correlation monotonically passes
through zero in the feromagnetic side, while the $\left\langle
s_{i}^{xy}s_{j}^{xy}\right\rangle$ correlation is maximal. The
remarkable quantum behavior of $\left\langle
s_{i}^{z}s_{j}^{z}\right\rangle$ around $R=0$ is discussed in Sec.V
below. It is seen in the figure that these changeovers are
increasingly sharp as temperature is decreased and, at zero
temperature, become discontinuous at $R=1$.  As seen in Fig.1(b), at
zero temperature, our calculated $\left\langle
s_{i}^{z}s_{j}^{z}\right\rangle$ and $\left\langle
s_{i}^{xy}s_{j}^{xy}\right\rangle$ correlations show very good
agreement with the known exact points \cite{Kato, Kitanine, Lieb,
Sato}.  Also, our results for $R>1$ fully overlap the exact results
of $\left\langle s_{i}^{z}s_{j}^{z}\right\rangle=0.25$ and
$\left\langle s_{i}^{xy}s_{j}^{xy}\right\rangle=0$ \cite{Takahashi}.
We also note that zero-temperature is the limit in which our
approximation is at its worst.

\section{Antiferromagnetic XXZ Chain}

\begin{figure}[h]
\centering
\includegraphics*[scale=1]{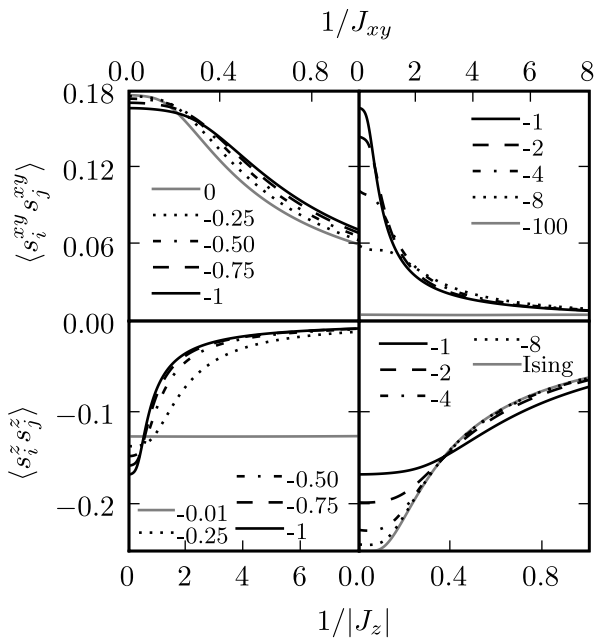}
\caption{Calculated nearest-neighbor spin-spin correlations
$\left\langle s_{i}^{xy}s_{j}^{xy}\right\rangle$ (upper panels) and
$\left\langle s_{i}^{z}s_{j}^{z}\right\rangle$ (lower panels) for
the antiferromagnetic XXZ chain, as a function of temperature, for
anisotropy coefficients $R=0, -0.25, -0.50, -0.75, -1, -2, -4, -8,
-\infty$ spanning the spin-liquid (left panels) and Isinglike (right
panels) regions.  Note that, in every one of the panels, the
correlation curves cross each other.  This remarkable quantum
phenomenon is discussed in the text.} \label{fig:2}
\end{figure}

\begin{figure}[h]
\centering
\includegraphics*[scale=0.9]{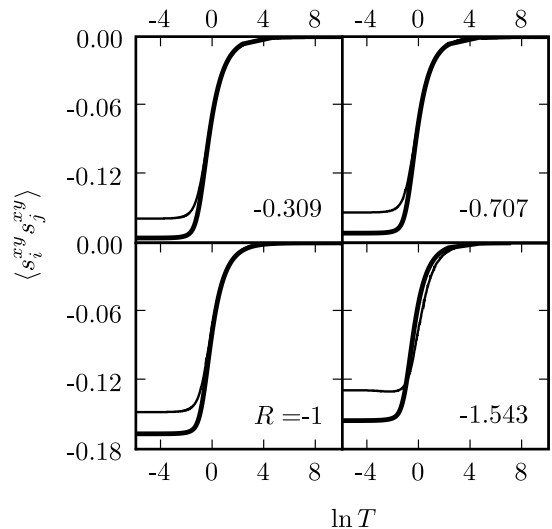}
\caption{Comparison of our results (thick lines) for the correlation
functions of the antiferromagnetic XXZ chain, with the
multiple-integral results of Ref.\cite{Bortz} (thin lines), for
various anisotropy coefficients $R$ spanning the spin-liquid and
Isinglike regions.} \label{fig:3}
\end{figure}

\begin{figure}[h]
\centering
\includegraphics*[scale=1]{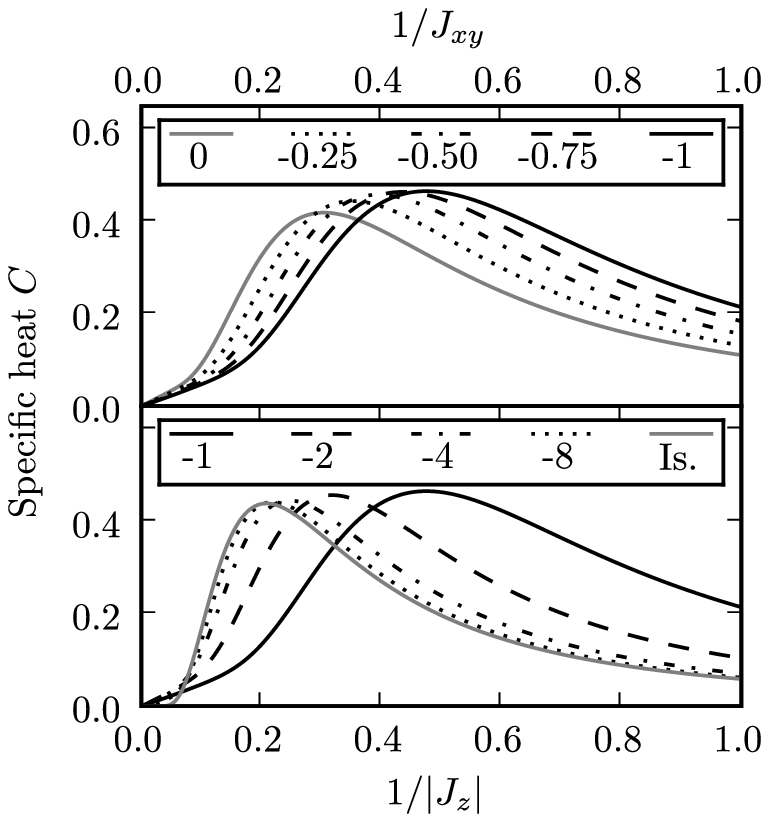}
\caption{Calculated specific heats $C$ of the antiferromagnetic XXZ
chain, as a function of temperature for anisotropy coefficients
$R=0, -0.25, -0.50, -0.75, -1, -2, -4, -8, -\infty$ spanning the
spin-liquid (upper panel) and Isinglike (lower panel) regions.}
\label{fig:4}
\end{figure}

\begin{figure}[h]
\centering
\includegraphics*[scale=1]{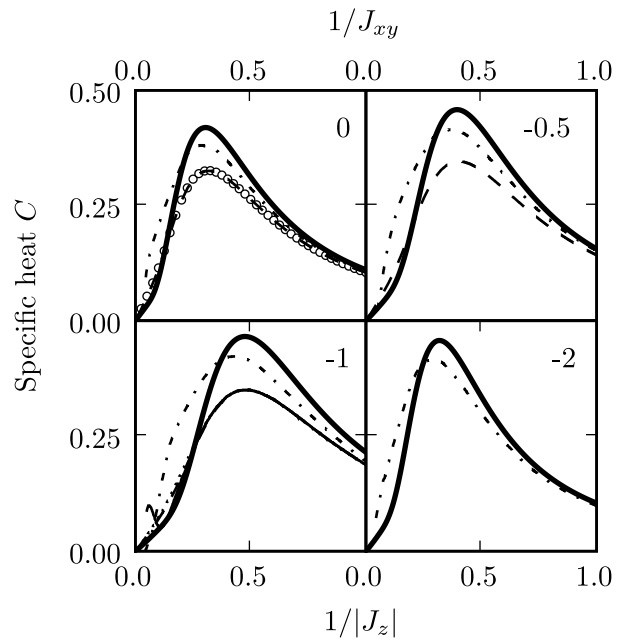}
\caption{Comparison of our antiferromagnetic specific heat results
(thick lines) with the results of Refs.\cite{Katsura} (open
circles), \cite{BF} (dotted), \cite{Narayanan} (thin lines),
\cite{Xi-Yao} (dash-dotted), and \cite{Fabricius, Klumper} (dashed),
for anisotropy coefficients $R=0,-0.5,-1,-2$ spanning the
spin-liquid and Isinglike regions.} \label{fig:6}
\end{figure}

For the antiferromagnetic XXZ chain, our calculated $\left\langle
s_{i}^{z}s_{j}^{z}\right\rangle$ and $\left\langle
s_{i}^{xy}s_{j}^{xy}\right\rangle$ nearest-neighbor spin-spin
correlations as a function of temperature are shown in
Fig.\ref{fig:2} for various anisotropy coefficients $R$.  We find
that when $\Jxy$ is the dominant interaction (spin liquid), the
correlations are weakly dependent on anisotropy $R$.  When $\Jz$ is
the dominant interaction (Isinglike), the correlations are weakly
dependent on anisotropy $R$ only at the higher temperatures. Our
results are compared with multiple-integral results \cite{Bortz} in
Fig.\ref{fig:3}.

In every one of the panels of Fig.\ref{fig:2}, the correlation
curves cross each other, revealing a remarkable quantum phenomenon.
In a classical system, the correlation between a given spin
component (\textit{e.g.}, $\left\langle
s_{i}^{xy}s_{j}^{xy}\right\rangle$) is expected to decrease when the
coupling of another spin component (\textit{e.g.}, $|J_z|$) is
increased.  It is found from the antiferromagnetic XXZ chain in
Fig.\ref{fig:2} that the opposite may occur in a quantum system: In
this figure, an increase in $J_{xy}$ causes an increase in
$|\left\langle s_{i}^zs_{j}^z\right\rangle|$ for $1/|J_z|
> 0.9$ and 0.4 in the spin-liquid and Isinglike regions respectively. Conversely, an increase in $|J_z|$ causes an increase in
$\left\langle s_{i}^{xy}s_{j}^{xy}\right\rangle$ for $1/J_{xy} >
0.4$ and 2.1 in the spin-liquid and Isinglike regions respectively.
This quantum effect can be called cross-component spin correlation.

\begin{figure}[h]
\centering
\includegraphics*[scale=1]{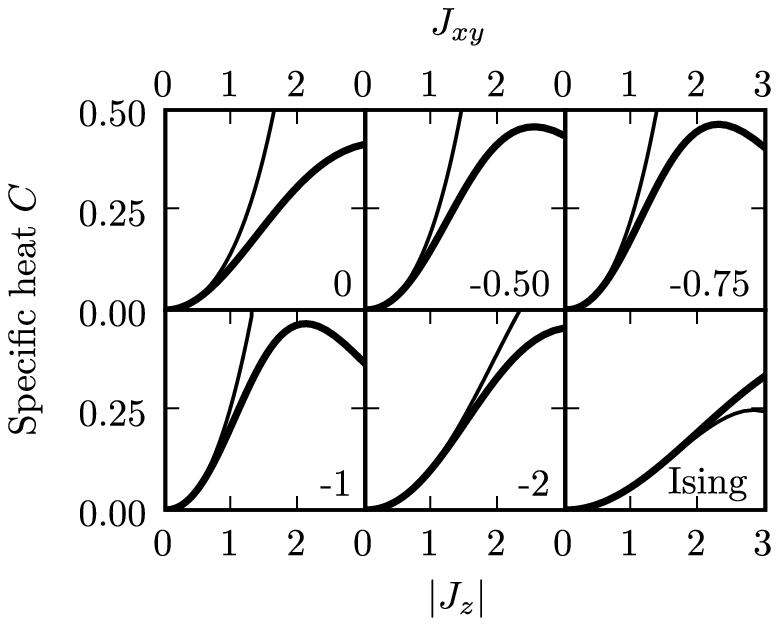}
\caption{Comparison of our antiferromagnetic specific heat results
(thick lines) with the high-temperature $J \rightarrow 0$ behaviors
(thin lines) obtained from series expansion in Ref.\cite{Rojas}, for
anisotropy coefficients $R=0,-0.50,-0.75,-1,-2,-\infty$ spanning the
spin-liquid and Isinglike regions.} \label{fig:6b}
\end{figure}

\begin{figure}[h]
\centering
\includegraphics*[scale=0.8]{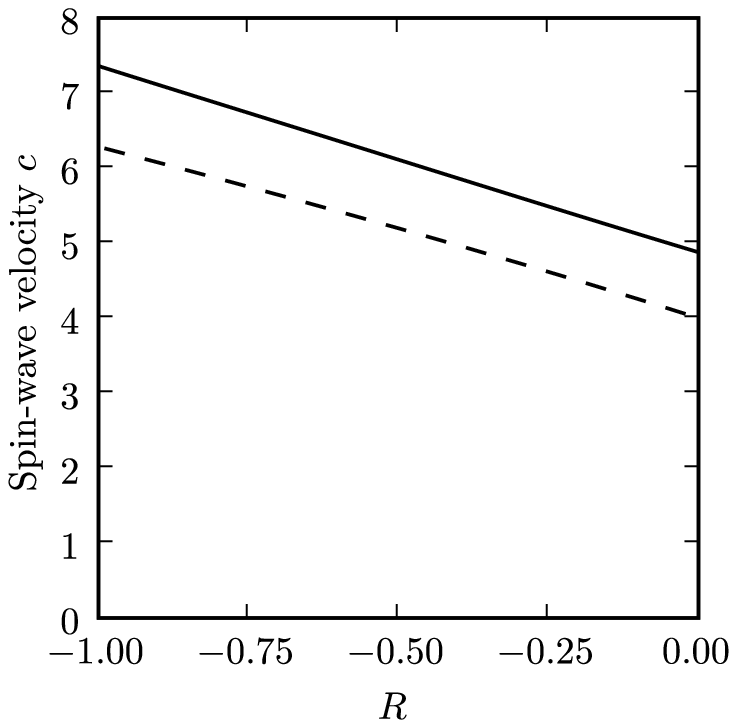}
\caption{Our calculated antiferromagnetic spin-wave velocity $c$
versus the anisotropy coefficient $R$. The dashed line, $2\pi \sin
(\gamma)/\gamma$ where $\gamma\equiv\cos^{-1}(-R)$, is the exact
result \cite{desCloizeaux}.} \label{fig:5}
\end{figure}

The antiferromagnetic specific heats calculated with
Eq.(\ref{eq:24}) are shown in Fig.\ref{fig:4} for various anisotropy
coefficients and compared, in Figs.\ref{fig:6}, \ref{fig:6b}, with
finite-lattice expansion \cite{BF, Narayanan}, quantum decimation
\cite{Xi-Yao}, transfer matrix \cite{Klumper}, high-temperature
series expansion \cite{Rojas} results and, for the $R=0$ case,
namely the XY model, with the exact result \cite{Katsura} $C=
(1/4\pi T)
\int^\pi_0\left(\cos\omega/\cosh\left(\frac{\cos\omega}{2T}\right)\right)^2d\omega$.
The $C(T)$ peak temperature is highest for the isotropic case
(Heisenberg) and decreases with anisotropy increasing in either
direction (towards Ising or XY). The peak value of $C(T)$ is only
weakly dependent on anisotropy, especially for the Isinglike
systems.  A strong contrast to this behavior will be seen, as
another quantum mechanical phenomenon, in the ferromagnetic XXZ
chain.

The linearity, at low temperatures, of the spin liquid ($|R|\leq1$)
specific heat with respect to temperature is expected on the basis
of spin-wave calculations for the antiferromagnetic XXZ model
\cite{Kubo, Kranendonk}. This linear form of $C(T)$ reflects the
linear energy-momentum dispersion of the low-lying excitations, the
magnons.  The low-temperature magnon dispersion relation is
$\hbar\omega=c k^n$, where $c$ is the spin-wave velocity and $n=1$
for the antiferromagnetic XXZ model in $d=1$ \cite{Takahashi}. The
internal energy, given by $U=(1/2\pi)\int_0^{\infty}dk
\hbar\omega(k)/(e^{\beta \hbar\omega(k)}-1)$, is dominated by the
magnons at low temperatures, yielding $U\sim T^2$ and $C\sim T$ for
$n=1$ in the dispersion relation. From this relation, our calculated
spin-wave velocity $c$ as a function of anisotropy $R$ is given in
Fig.\ref{fig:5} and compares well with the also shown exact result
\cite{desCloizeaux}.  A simultaneous fit to the dispersion relation
exponent $n$, expected to be 1, yields $1.00\pm0.02$. However, for
the Isinglike $-R>1$, the unexpected linearity instead of an
exponential form caused by a gap in the excitation spectrum, points
to the approximate nature of our renormalization-group calculation.
The correct exponential form is obtained in the large $-R$ limit,
where the renormalization-group calculation becomes exact.

\begin{figure}[h]
\centering
\includegraphics*[scale=1]{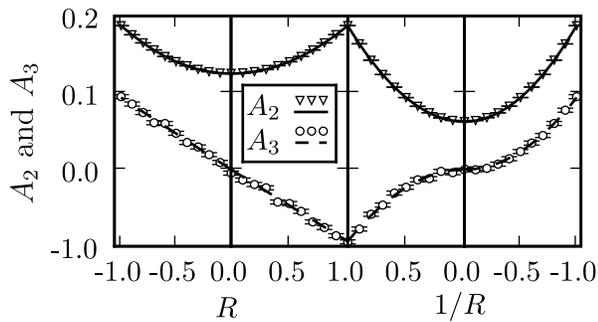}
\caption{Comparison of our results with the high-temperature
expansion of Ref.~\cite{Rojas} for all regions: antiferromagnetic
(outer panels) and ferromagnetic (inner panels), spin-liquid (left
panels) and Isinglike (right panels). Triangles and circles denote
our results, while solid and dashed lines denote the results of
Ref.\cite{Rojas} for $A_2$ and $A_3$, respectively. The error bars,
due to the statistical fitting procedure of the coefficients $A_2$
and $A_3$, have half-heights of $1.7\times10^{-4}$ and
$2.6\times10^{-3}$ respectively.} \label{fig:7}
\end{figure}

Rojas et al. \cite{Rojas} have obtained the high-temperature
expansion of the free energy of the XXZ chain to order $\beta^3$,
where $\beta$ is the inverse temperature. The specific heat from
this expansion is

\begin{eqnarray}
C=\frac{2+R^2}{16}\Jxy^2-\frac{3R}{32}\Jxy^3+\frac{6-8R^2-R^4}{256}\Jxy^4.
\label{eq:a03}
\end{eqnarray}
This high-temperature specific heat result is also compared with our
results, in Fig.\ref{fig:6b}, and very good agreement is seen. In
fact, when in the high-temperature region of $0<\beta<0.1$, we fit
our numerical results for $C(\beta)$ to the fourth degree polynomial
$C=\Sigma_{i=0}^4A_i\beta^i$, and we do find (1) the vanishing
$A_0<10^{-5}$ and $A_1<10^{-7}$ for all $R$ and (2) the comparison
in Fig.\ref{fig:7} between our results for $A_2$ and $A_3$ and those
of Eq.(\ref{eq:a03}) from Ref.\cite{Rojas}, thus obtaining excellent
agreement for all regions of the model.

\begin{figure}[h]
\centering
\includegraphics*[scale=1]{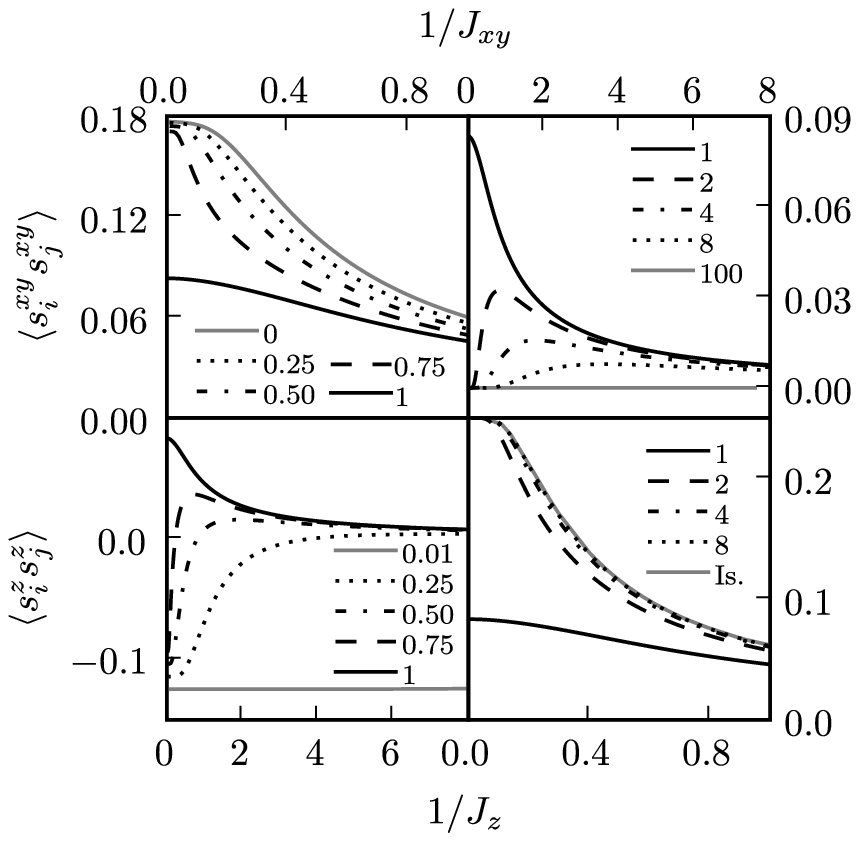}
\caption{Calculated nearest-neighbor spin-spin correlations
$\left\langle s_{i}^{xy}s_{j}^{xy}\right\rangle$ (upper panels) and
$\left\langle s_{i}^{z}s_{j}^{z}\right\rangle$ (lower panels) for
the ferromagnetic XXZ chain, as a function of temperature, for
anisotropy coefficients $R=0,0.25,0.50,0.75,1,2, 4,8,\infty$
spanning the spin-liquid (left panels) and Isinglike (right panels)
regions.} \label{fig:17}
\end{figure}

\begin{figure}[h]
\centering
\includegraphics*[scale=1]{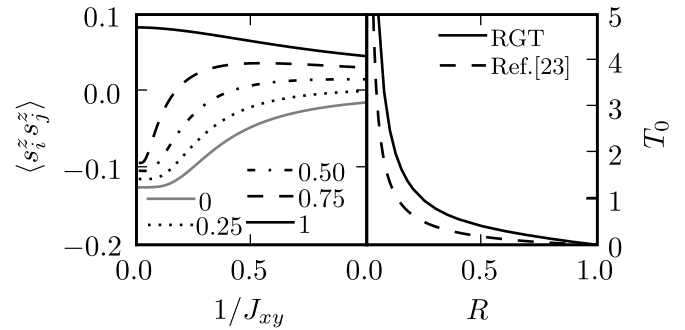}
\caption{Left panel:  Calculated nearest-neighbor spin-spin
correlations $\left\langle s_{i}^{z}s_{j}^{z}\right\rangle$ for the
ferromagnetic XXZ chain, as a function of temperature $1/J_{xy}$ in
the spin liquid, for anisotropy coefficients $R=0,0.25,0.50,0.75,1$.
Right panel:  The sign-reversal temperature $T_0$ of the
nearest-neighbor correlation $\langle s_{i}^{z}s_{j}^{z}\rangle$:
our results (full curve) and the analytical result from the quantum
transfer matrix method (dashed) \cite{Fabricius}.} \label{fig:a02}
\end{figure}

\begin{figure}[h]
\centering
\includegraphics*[scale=0.95]{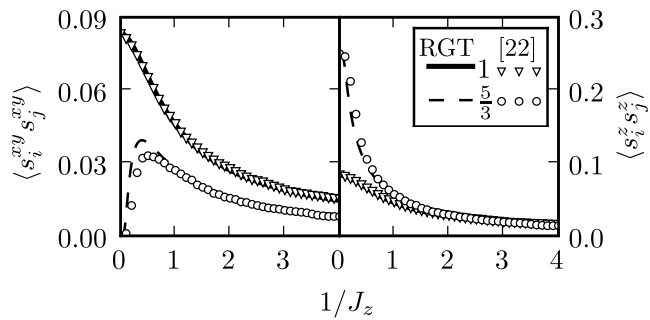}
\caption{Comparison of our ferromagnetic $R=1,\frac{5}{3}$ results
with Green's function calculations \cite{Zhang}~.} \label{figb:17}
\end{figure}

\begin{figure}[h]
\centering
\includegraphics*[scale=1]{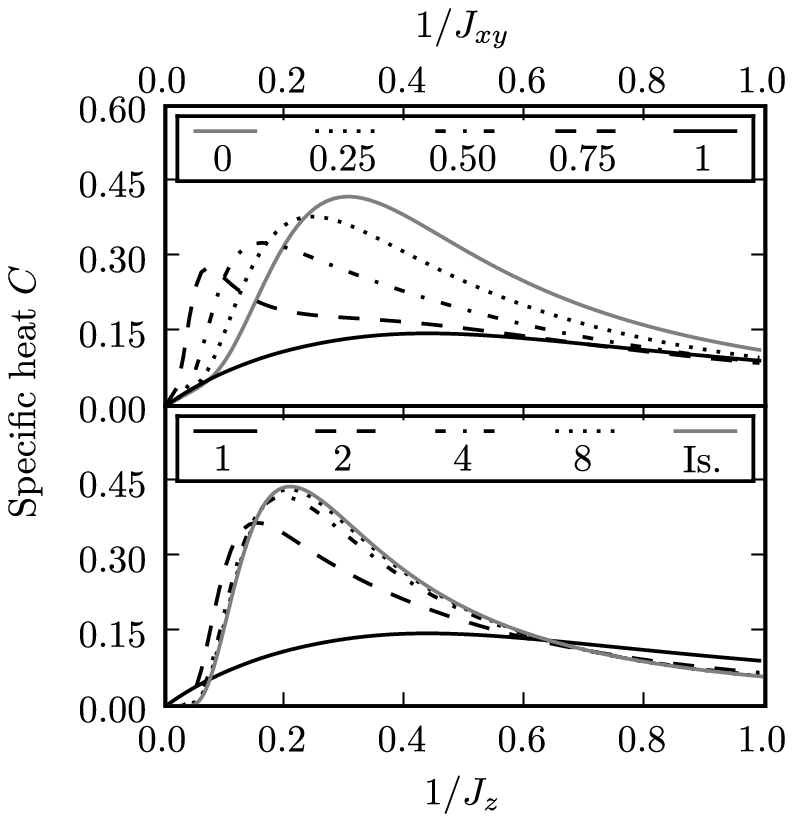}
\caption{Calculated specific heats $C$ of the ferromagnetic XXZ
chain, as a function of temperature for anisotropy coefficients
$R=0,0.25,0.50,0.75,1,2,4,8,\infty$ spanning the spin-liquid (upper
panel) and Isinglike (lower panel) regions.} \label{fig:22}
\end{figure}

\begin{figure}[h]
\centering
\includegraphics*[scale=1]{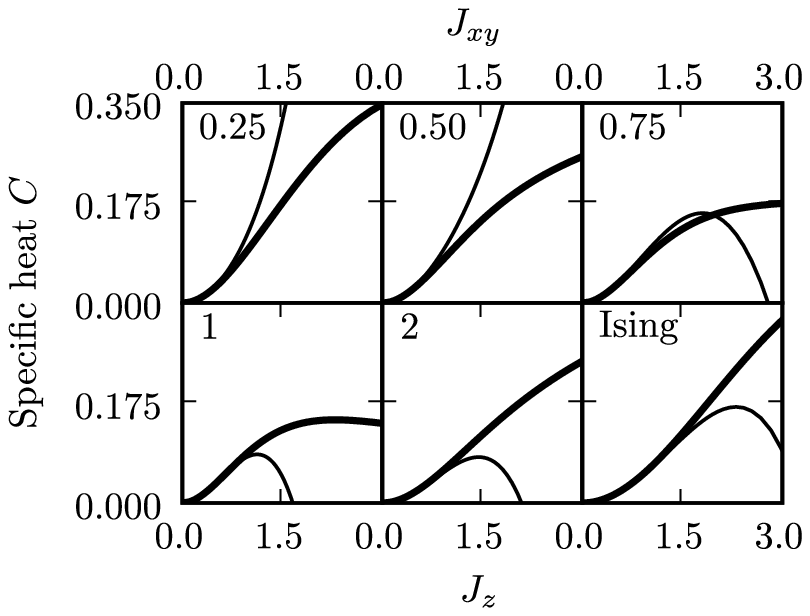}
\caption{Comparison of our ferromagnetic specific heat results
(thick lines) with the high-temperature $J \rightarrow 0$ behaviors
(thin lines) obtained from series expansion \cite{Rojas}, for
anisotropy coefficients $R=0.25,0.50,0.75,1,2,\infty$ spanning the
spin-liquid and Isinglike regions.} \label{fig:a21}
\end{figure}

\begin{figure}[h]
\centering
\includegraphics*[scale=1]{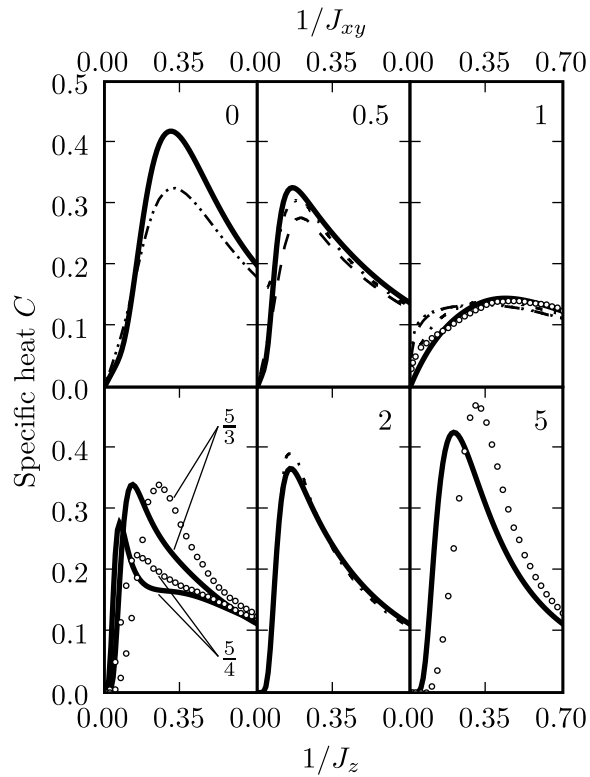}
\caption{Comparison of our ferromagnetic specific heat results
(thick lines) with the results of Refs.\cite{Katsura}
(dash-double-dotted), \cite{BF} (dotted), \cite{Xi-Yao}
(dash-dotted), \cite{Zhang} (open circles), and \cite{Fabricius,
Klumper} (dashed), for anisotropy coefficients
$R=0,0.5,1,\frac{5}{4},\frac{5}{3},2,5$ spanning the spin-liquid and
Isinglike regions.} \label{fig:a08}
\end{figure}

\begin{figure}[h]
\includegraphics*[scale=0.8]{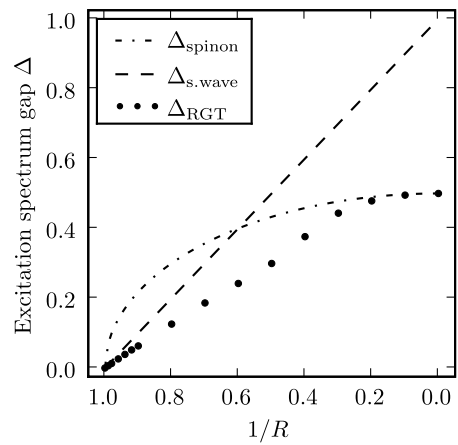}
\caption{The calculated excitation spectrum gap $\Delta$ versus
anisotropy.} \label{fig:24b}
\end{figure}

\begin{table}[h]
\begin{tabular}{|c|c|c|}
\hline
\parbox{1.1in}{\centering Zero-temperature correlations of the spin-$\frac{1}{2}$ XY chain} & \parbox{0.75in}{\centering Exact values from Ref.~\cite{Lieb}} & \parbox{0.4in}{\centering Our RGT results} \\
\hline
$\langle s_i^{xy}s_j^{xy}\rangle$ & $\:\:~0.15915$ & $\:\:~0.17678$ \\
\hline
$\langle s_i^{z}s_j^{z}\rangle$ & $-0.10132$ & $-0.12500$\\
\hline
\end{tabular}
\caption{Zero-temperature nearest-neighbor correlations of the
spin-$\frac{1}{2}$ XY chain.} \label{tab:7}
\end{table}

\section{Ferromagnetic XXZ Chain}

For the ferromagnetic (\textit{i.e.}, $R>0$) systems in
Fig.\ref{fig:1}, the $\left\langle s_i^{z}s_j^{z}\right\rangle$
expectation value becomes rapidly negative at lower temperatures for
$R<1$, even though for $R\geq 0$ all couplings in the Hamiltonian
are ferromagnetic. This is actually a real physical effect, not a
numerical anomaly. In fact, we know the spin-spin correlations for
the ground state of the one-dimensional XY model (the $R=0$ case of
our Hamiltonian), and we can compare our low-temperature results
with these exact values. The ground-state properties of the
spin-$\frac{1}{2}$ XY model are studied by making a Jordan-Wigner
transformation, yielding a theory of non-interacting spinless
fermions.  Analysis of this theory yields the exact zero-temperature
nearest-neighbor spin-spin correlations~\cite{Lieb} shown in Table
\ref{tab:7}. Our renormalization-group results in the
zero-temperature limit, also shown in this table, compare quite well
with the exact results, as with the other exact points in Fig.1(b),
although in the worst region for our approximation. Finally, by
continuity, it is reasonable that for a range of $R$ positive but
less than one, the $z$ component correlation function is as we find,
intriguingly but correctly negative at low temperatures. Thus, the
interaction $s_{i}^{x}s_{j}^{x}+s_{i}^{y}s_{j}^{y}$ (irrespective of
its sign, due to the symmetry mentioned at the end of Sec.IIB)
induces an antiferromagnetic correlation in the $s_{i}^{z}$
component, competing with the $s_{i}^{z}s_{j}^{z}$ interaction when
the latter is ferromagnetic.

For finite temperatures, our calculated nearest-neighbor spin-spin
correlations are shown in Figs.\ref{fig:17}, \ref{fig:a02}, for
different values of $R$. These results are compared with Green's
function calculations \cite{Zhang} in Fig.\ref{figb:17}. As expected
from the discussion at the beginning of this section, in the
spin-liquid region, the correlation $\langle
s_{i}^{z}s_{j}^{z}\rangle$ is negative at low temperatures. Thus, a
competition occurs in the correlation $\langle
s_{i}^{z}s_{j}^{z}\rangle$ between the XY-induced antiferromagnetism
and the ferromagnetism due to the direct coupling between the
$s^{z}$ spin components.  In fact, the reinforcement of
antiferromagnetic correlations of $\langle
s_{i}^{z}s_{j}^{z}\rangle$ by increasing $J_{xy}$ (and also its
converse) was seen in the antiferromagnetic XXZ chain discussed in
the previous section.  Thus, we see that whereas this
cross-component effect is dominant at low temperatures in the
ferromagnetic XXZ chain, it is seen at higher temperatures in the
antiferromagnetic XXZ  chain and, in-between, throughout the
temperature range in the XY chain.

In the ferromagnetic XXZ chain, as a consequence of the competition
mentioned above, a sign reversal in $\langle
s_{i}^{z}s_{j}^{z}\rangle$ occurs from negative to positive
correlation, at temperatures $J_{xy}^{-1} =
T_0(R)$.\cite{Schindelin} At this temperature, by cancelation of the
competing effects, the nearest-neighbor correlation $\langle
s_{i}^{z}s_{j}^{z}\rangle$ is zero. Our calculated $T_0(R)$ curve is
shown in Fig.\ref{fig:a02}, and has very good agreement with the
exact result $T_0=({\sqrt{3}\sin\gamma}/4\gamma)
\tan[\pi(\pi-\gamma)/2\gamma]$ where $\gamma\equiv\cos^{-1}(-R)$
\cite{Fabricius}.

The calculated ferromagnetic specific heats are shown in
Fig.\ref{fig:22} for various anisotropy coefficients and compared,
in Figs.\ref{fig:a21}, \ref{fig:a08}, with finite-lattice expansion
\cite{BF}, quantum decimation \cite{Xi-Yao}, decoupled Green's
functions \cite{Zhang}, transfer matrix \cite{Fabricius, Klumper},
high-temperature series expansion \cite{Rojas} results and, for the
$R=0$ case, namely the XY model, with the exact result
\cite{Katsura} $C= (1/4\pi T)
\int^\pi_0\left(\cos\omega/\cosh\left(\frac{\cos\omega}{2T}\right)\right)^2d\omega$.
In sharp contrast to the antiferromagnetic case in Sec.IV, the peak
$C(T)$ temperature is highest for the most anisotropic cases (XY or
Ising) and decreases with anisotropy decreasing from either
direction (towards Heisenberg).  In the same contrast, the peak
value of $C(T)$ is dependent on anisotropy, decreasing, eventually
to a flat curve, as anisotropy is decreased.  This contrast between
the ferromagnetic and antiferromagnetic systems is a purely quantum
phenomenon.  Specifically, the marked contrast between the specific
heats of the isotropic antiferromagnetic and ferromagnetic systems,
seen in the full curves of Figs.\ref{fig:4} and \ref{fig:22}
respectively, translates into the different critical temperatures of
the respective three-dimensional systems.\cite{Rushbrooke, Oitmaa,
Falicov, Hinczewski} Classical ferromagnetic and antiferromagnetic
systems are, on the other hand, identically mapped onto each other.

The low-temperature specifics heats are discussed in detail and
compared to other results in Sec.VI.

\begin{figure}[h]
\centering
\includegraphics*[scale=1]{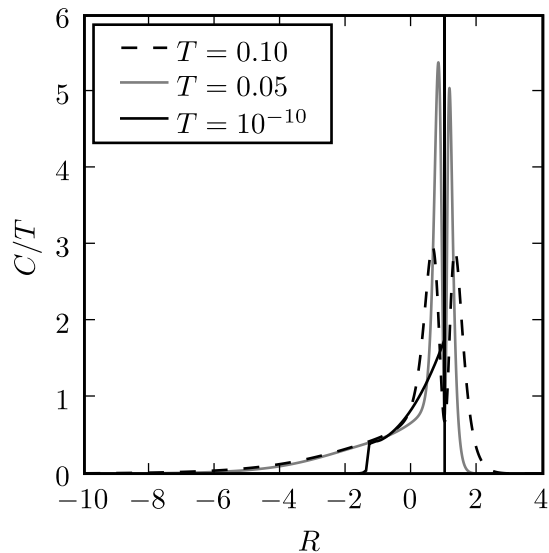}
\caption{Calculated specific heat coefficient $C/T$ as a function of
anisotropy $R$, for $T=0.10, 0.05, 10^{-10}$\:.} \label{fig:26}
\end{figure}

\begin{figure}[h]
\centering
\includegraphics*[scale=1]{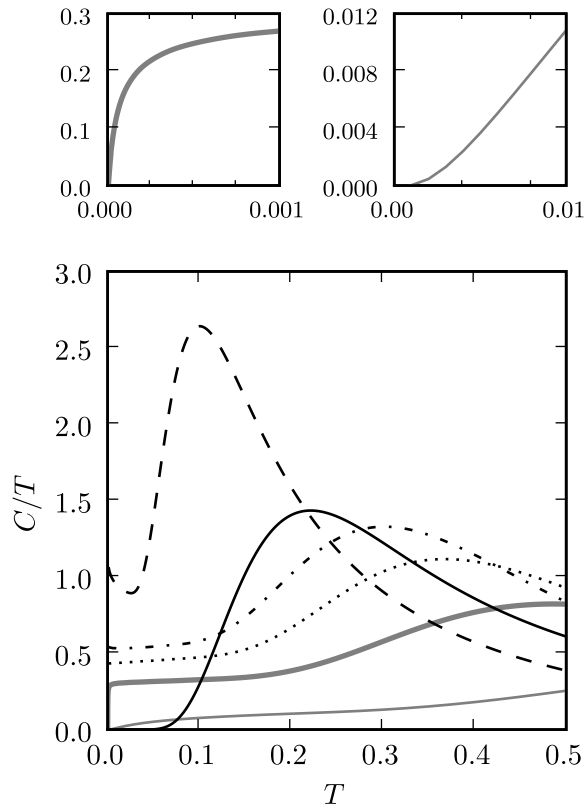}
\caption{Calculated specific heat coefficient $C/T$ as a function of
temperature for anisotropy coefficient $R = -5$ (thin grey), $R =
-2$ (thick grey), -1 (dotted), -0.5 (dash-dotted), 0.5 (dashed), and
2 (thin black).} \label{fig:27}
\end{figure}

\begin{figure}[b]
\centering
\includegraphics*[scale=1]{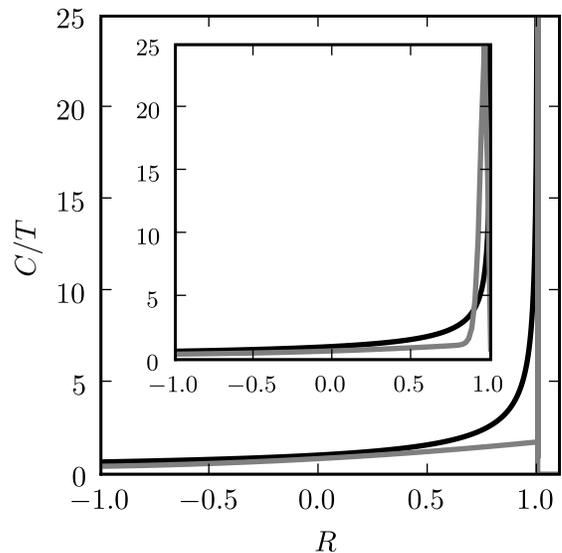}
\caption{Calculated specific heat coefficient $C/T$ as a function of
anisotropy coefficient $R$ in the spin-liquid region, $-1\leq
R\leq1$, at constant temperature $T=10^{-10}$\:. Our
renormalization-group result (grey curve) is compared to the
zero-temperature Bethe-Ansatz result (black curve).  Inset: our
calculation (grey curve) at constant $T=10^{-2}$ is again compared
to the zero-temperature Bethe-Ansatz result (black curve).}
\label{fig:28}
\end{figure}

\section{Low-Temperature Specific Heats}

Properties of the low-temperature specific heat of the ferromagnetic
XXZ chain have been derived from the thermodynamic Bethe-ansatz
equations \cite{Takahashi}.  For anisotropy coefficient $|R|\leq1$,
the model is gapless \cite{Haldane1980,Haldane1982} and, except at
$R=1$, the specific heat is linear in $T = J_{xy}^{-1}$ in the
zero-temperature limit, $C/T=2\gamma/(3\sin\gamma)$ where again
$\gamma\equiv\cos^{-1}(-R)$. Note that this result contradicts the
spin-wave theory prediction of $C\sim T^{1/2}$ for the ferromagnetic
chain ($n=2$ for the ferromagnetic magnon dispersion relation of the
kind given above in Sec.IV). The spin-wave result is valid only for
$R=1$, the isotropic Heisenberg case. From the expression given
above, we see that $C/T$ diverges as $R\rightarrow1^-$, and at
exactly $R=1$ it has been shown that $C\sim T^{1/2}$
\cite{Takahashi}.

In the Isinglike region $R>1$, the system exhibits a gap in its
excitation spectrum and the specific heat behaves as $C\sim
T^{-a}\:\exp(-\Delta/T)$, with $\Delta$ being the excitation
spectrum gap \cite{Haldane1980,Haldane1982,Takahashi}.  There exist
two gaps for the energy, called the spinon gap and the spin-wave
gap, given by $\Delta_{spinon}=\frac{1}{2}\sqrt{1-R^{-2}}$ and
$\Delta_{spinwave}=1-R^{-1}$. These are the minimal energies of
elementary excitations~\cite{Takahashi, Johnson}. A crossover
between them occurs at $R=\frac{5}{3}$: below this value, the spinon
gap is lower, while above this value the spin-wave gap is lower. We
have double-fitted our calculated specific heats with respect to the
gap $\Delta$ and the leading exponent $a$, for the entire range of
anisotropy $R$ between $0 < R^{-1} < 1$ (Fig.\ref{fig:24b}).  Our
calculated gap $\Delta$ behaves linearly in $R^{-1}$ for $R^{-1}$
close to 1, and crosses over to 1/2 at $R^{-1}=0$, as expected.  We
also obtain the exponent $a = 1.99 \pm 0.02$ in the Ising limit
$R^{-1} \leq 0.2$ and $a = 1.52 \pm 0.10$ in the Heisenberg limit
$R^{-1} \geq 0.9$.  These exponent values are respectively expected
to be 2 and 1.5 \cite{Johnsonb, Johnson}.

We now turn to the discussion of our specific heat results for the
entire ferromagnetic and antiferromagnetic ranges. Our calculated
$C/T$ curves are plotted as a function of anisotropy and temperature
in Figs.\ref{fig:26} and \ref{fig:27} respectively. We discuss each
region of the anisotropy $R$ separately:

\noindent (i) $R>1$~: The specific heat coefficient $C/T$ vanishes
in the $T\rightarrow0$ limit and has the expected exponential form
as discussed above in this section.  The spin-wave to spinon
excitation gap crossover is obtained.

\noindent (ii) $R\approx1$~: The double-peak structure of $C/T$ in
Fig.\ref{fig:26} is centered at $R=1$. As temperature goes to zero,
the peaks narrow and diverge.

\noindent (iii) $-1\leq R<1$~:  The specific heat coefficient is
$C/T=2\gamma/(3\sin\gamma)$ in this region \cite{Haldane1980,
Takahashi}, and our calculated specific heat is indeed linear at low
temperatures. The $C/T$ curves for $R=-1,-0.5,0.5$ in
Fig.\ref{fig:27} all extrapolate to nonzero limits at $T=0$.  The
spin-wave dispersion relation exponent and velocity, for the
antiferromagnetic system, is correctly obtained for the isotropic
case and for all anisotropies, as seen in Fig.\ref{fig:5}.
Fig.\ref{fig:28} directly compares $C/T=2\gamma/(3\sin\gamma)$ with
our results: The curves have the same basic form, gradually rising
from $R=-1$, with a sharp divergence as $R$ nears $1$.  At $R=1^+$,
we expect $C/T=0$. Our $T=10^{-10}$ curve diverges at $R=1$ and
indeed returns to zero at $R=1.0000001$.

\noindent (iv) $R<-1$~:  We expect a vanishing $C/T$, which we do
find as seen in Fig.\ref{fig:26} and in the insets of
Fig.\ref{fig:27}.  The exponential behavior of the specific heat is
clearly seen in the Ising limit.

\section{Conclusion}

A detailed global renormalization-group solution of the XXZ
Heisenberg chain, for all temperatures and anisotropies, for both
ferromagnetic and antiferromagnetic couplings, has been obtained. In
the spin-liquid region, the linear low-temperature specific heat
and, for the antiferromagnetic chain, the spin-wave dispersion
relation exponent $n$ and velocity $c$ have been obtained. In the
Isinglike region, the spin-wave to spinon crossover of the
excitation spectrum gap of the ferromagnetic chain has been obtained
from the exponential specific heat, as well as the correct leading
algebraic behaviors in the Heisenberg and Ising limits.  Purely
quantum mechanical effects have been seen:  We find that the $xy$
correlations and the antiferromagnetic $z$ correlations mutually
reinforce each other, for different ranges of temperatures and
anisotropies, in ferromagnetic, antiferromagnetic, and XY systems.
The behaviors, with respect to anisotropy, of the specific heat peak
values and locations are opposite in the ferromagnetic and
antiferromagnetic systems. The sharp contrast found in the specific
heats of the isotropic ferromagnetic and antiferromagnetic systems
is a harbinger of the different critical temperatures in the
respective three-dimensional systems. When compared with existing
calculations in the various regions of the global model, good
quantitative agreement is seen. Even at zero temperature, where our
approximation is at its worst, good quantitative agreement is seen
with exact data points for the correlation functions
(Fig.\ref{fig:1}(b)), which we extend to all values of the
anisotropy. Finally, the relative ease with which the Suzuki-Takano
decimation procedure is globally and quantitatively implemented
should be noted.

\begin{acknowledgments}
This research was supported by the Scientific and Technological
Research Council (T\"UB\.ITAK) and by the Academy of Sciences of
Turkey.  One of us (O.S.S.) gratefully acknowledges a scholarship
from the Turkish Scientific and Technological Research Council -
Scientist Training Group (T\"UB\.ITAK-BAYG).
\end{acknowledgments}

\end{document}